\documentclass[letterpaper]{article} 
\usepackage{aaai25}  
\usepackage{times}  
\usepackage{helvet}  
\usepackage{courier}  
\usepackage[hyphens]{url}  
\usepackage{graphicx} 
\usepackage{natbib}  
\usepackage{caption} 
\usepackage{algorithm}
\usepackage{algorithmic}

\usepackage{newfloat}
\usepackage{listings}
\DeclareCaptionStyle{ruled}{labelfont=normalfont,labelsep=colon,strut=off} 
\floatstyle{ruled}
\newfloat{listing}{tb}{lst}{}
\floatname{listing}{Listing}
\usepackage{pifont}
\usepackage{graphicx}
\usepackage{amsmath}
\usepackage{booktabs}
\usepackage{tabularx}
\usepackage{multirow}
\usepackage{url}

\usepackage{booktabs}
\usepackage{array}
\usepackage{multirow}
\usepackage{makecell}
\usepackage{enumitem}
\usepackage{amssymb}
\usepackage{amssymb}

\title{ConDSeg: A General Medical Image Segmentation Framework via Contrast-Driven Feature Enhancement}
\author{
    Mengqi Lei\textsuperscript{\rm 1},
    Haochen Wu\textsuperscript{\rm 1},
    Xinhua Lv\textsuperscript{\rm 1},
    Xin Wang\textsuperscript{\rm 2}
}
\affiliations{
    \textsuperscript{\rm 1}China University of Geosciences, Wuhan 430074, China\\
    \textsuperscript{\rm 2}Baidu Inc, Beijing, China\\
    mengqi-lei@163.com, hcwu@cug.edu.cn, xinhualv@cug.edu.cn, wangxin105@baidu.com
    
}

\usepackage{bibentry}

\begin{document}

\maketitle

\begin{abstract}
Medical image segmentation plays an important role in clinical decision making, treatment planning, and disease tracking. However, it still faces two major challenges. On the one hand, there is often a ``soft boundary'' between foreground and background in medical images, with poor illumination and low contrast further reducing the distinguishability of foreground and background within the image. On the other hand, co-occurrence phenomena are widespread in medical images, and learning these features is misleading to the model's judgment. To address these challenges, we propose a general framework called Contrast-Driven Medical Image Segmentation (ConDSeg). First, we develop a contrastive training strategy called Consistency Reinforcement. It is designed to improve the encoder's robustness in various illumination and contrast scenarios, enabling the model to extract high-quality features even in adverse environments. Second, we introduce a Semantic Information Decoupling module, which is able to decouple features from the encoder into foreground, background, and uncertainty regions, gradually acquiring the ability to reduce uncertainty during training. The Contrast-Driven Feature Aggregation module then contrasts the foreground and background features to guide multi-level feature fusion and key feature enhancement, further distinguishing the entities to be segmented. We also propose a Size-Aware Decoder to solve the scale singularity of the decoder. It accurately locate entities of different sizes in the image, thus avoiding erroneous learning of co-occurrence features. Extensive experiments on five medical image datasets across three scenarios demonstrate the state-of-the-art performance of our method, proving its advanced nature and general applicability to various medical image segmentation scenarios. 
Our released code is available at \url{https://github.com/Mengqi-Lei/ConDSeg}.
\end{abstract}
\section{Introduction}
Medical image segmentation is a cornerstone of the medical imaging field, providing indispensable support for clinical decision-making, treatment planning, and disease monitoring \cite{xun2022survey,qureshi2023survey2}. 
In recent years, deep learning methods \cite{xi2024evaluating} have shown great potential in medical image segmentation \cite{liu2021deepl2,malhotra2022deepl}, such as U-Net \cite{ronneberger2015unet}, U-Net++ \cite{zhou2018unet++}, PraNet \cite{fan2020pranet}, TGANet \cite{tomar2022tganet}, CASF-Net \cite{zheng2023casf-net}, continuously improving segmentation performance.

\begin{figure}
    \centering
    \includegraphics[width=0.95\linewidth]{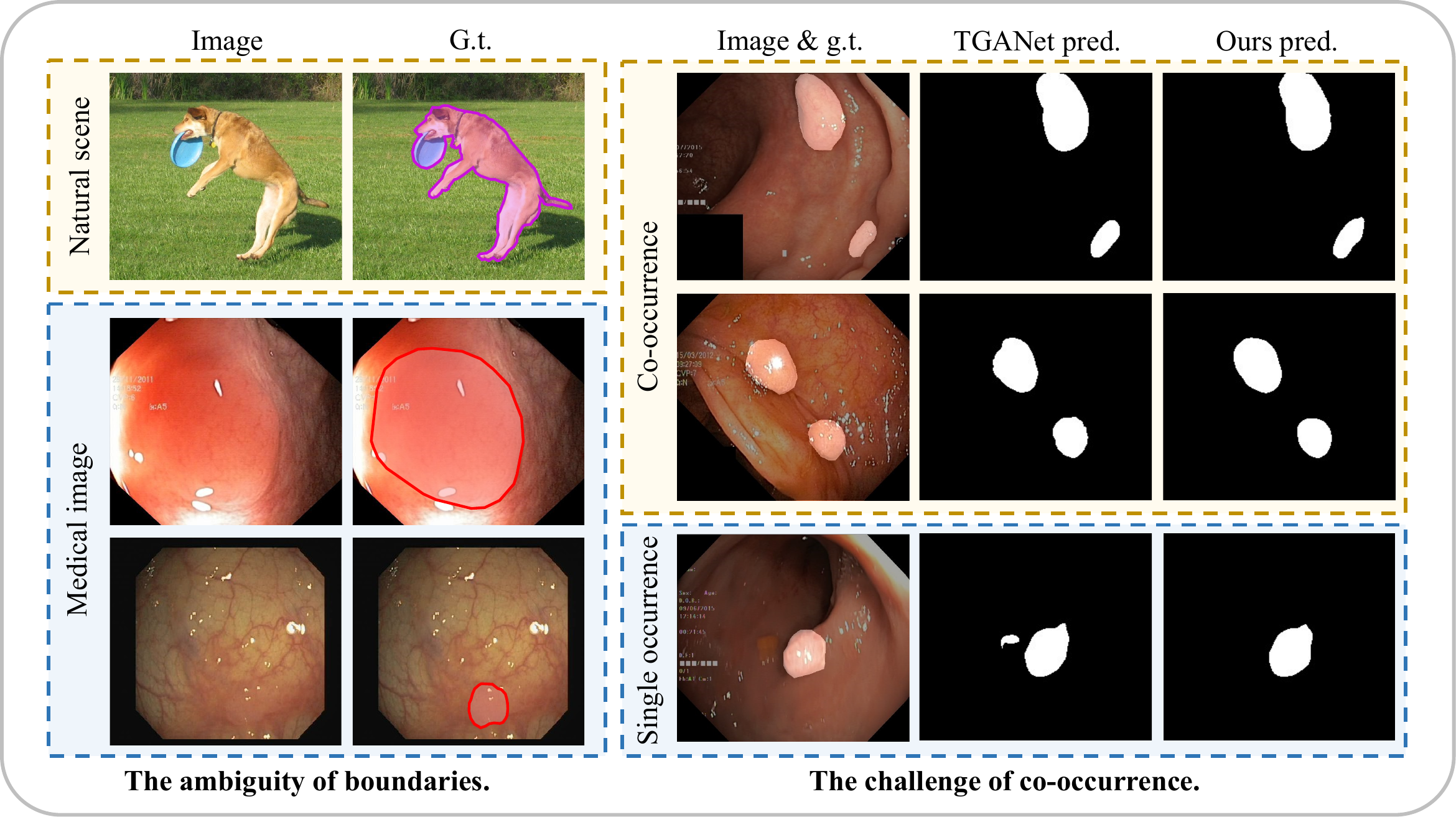}
    \caption{Major challenges in medical image segmentation.}
    \label{fig:challenges}
    \vspace{-10pt}
\end{figure}

In the field of medical images, different scenarios of tasks use images with different modalities, such as endoscopy images \cite{jha2020kvasir-seg,jha2021sessile}, dermatoscopy images \cite{gutman2016isic-2016,codella2018isic-2017}, digitally scanned whole slide images \cite{sirinukunwattana2017glas}, etc. Although existing deep learning methods have made great breakthroughs in medical image segmentation, achieving accurate segmentation is still a challenge, which comes from two main aspects: ambiguous boundaries and co-occurrence phenomena, which are visualized in Figure \ref{fig:challenges}.
On the one hand, compared with natural images where the boundary between foreground and background is clear, in medical images, the foreground and background are often a kind of blurred ``soft boundary'' \cite{xie2023boundary, lei2024epps}. This blurring is mainly due to the existence of transition areas between pathological tissue and surrounding normal tissue, making the boundary hard to define. Moreover, in many cases, medical images exhibit poor light and low contrast \cite{cossio2023low_light,9363892low_contrast}, which further blurs the boundary between pathological and normal tissues, adding to the difficulty of distinguishing the boundary. 
On the other hand, co-occurrence is widely present in medical images \cite{chen2022co-occurrence}. For example, in endoscopy polyp images, small polyps tend to co-occur with polyps of similar size. 
This leads the model to easily learn certain co-occurrence features that are not related to the polyps themselves. And when the pathological tissue appears alone, the model often fails to predict accurately. Thus, We propose a general framework called Contrast-Driven Medical Image Segmentation (ConDSeg) to overcome these challenges.

To address the challenges posed by ambiguous boundaries, we first introduce a preliminary training strategy named Consistency Reinforcement (CR). 
Specifically, we feed the encoder with the original images and the strongly augmented images, and use the output of the encoder to predict the masks respectively.
By maximizing the consistency between these mask pairs, we enhance the Encoder's robustness to various scenes involving different lighting conditions, colors, etc., enabling it to extract high-quality features even in adverse environments. 
Subsequently, we propose the Semantic Information Decoupling (SID) module, which decouples the feature map from the Encoder into three distinct ones: foreground, background and regions of uncertainty. 
During the training phase, we promote the reduction of the uncertainty regions by enforcing their complementarity and the precision of the foreground and background masks through our elaborately designed loss functions.

\begin{figure*}[t]
    \centering
    \includegraphics[width=0.95\linewidth]{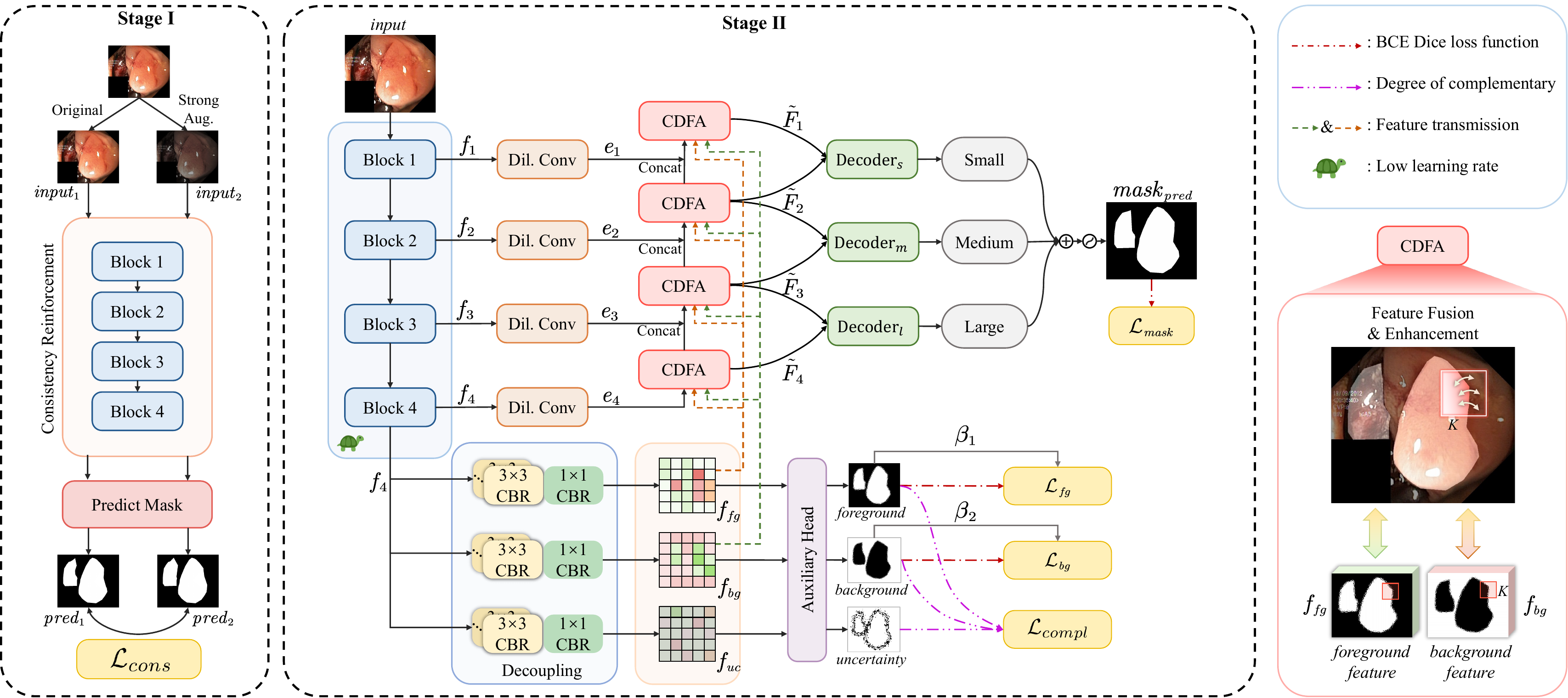}
    \caption{Overall framework of the proposed ConDSeg.}
    \label{fig:framework}
\end{figure*}

To further emphasize the distinction between foreground and background and to address the challenges brought by co-occurrence phenomena, we propose the Contrast-Driven Feature Aggregation (CDFA) module. CDFA receives foreground and background features from SID, guiding the multi-level feature fusion and the enhancement of key features under their contrastive information. Further, We propose Size-Aware Decoder to solve the problem of scale singularity of decoders \cite{cao2022swin-unet, wang2023xboundformer} when predicting masks.
The SA-Decoders responsible for different sizes individually receive feature maps from different levels to predict entities of various sizes, thereby facilitating the model's ability to distinguish between different entities within the same image and preventing the erroneous learning of co-occurrence features. 
Overall, the main contributions of this paper are as follows:
\begin{itemize}[topsep=1pt,leftmargin=*]
    \item Our proposed CR preliminary training strategy can effectively improve the robustness of the Encoder to harsh environments and thus extract high-quality features. On the other hand, SID can decouple feature maps into foreground, background, and regions of uncertainty, and it learns to reduce uncertainty during the training process via specially designed loss functions.
    \item Our proposed CDFA guides the fusion and enhancement of multilevel features via the contrastive features extracted by SID. The SA-Decoder is designed to better differentiate between different entities within the image and make separate predictions for entities of various sizes, overcoming the interference of co-occurrence features. 
    \item We have conducted extensive comparative and ablation experiments on five medical image datasets, covering tasks under three modalities. Our ConDSeg achieves state-of-the-art performance across all datasets. 
\end{itemize}

\vspace{-2mm}
\section{Related Works}

Automatic medical image segmentation has been one of the research hotspots in the field of medical images \cite{xun2022survey,wang2022survey3,qureshi2023survey2}. Over the past decade, with the rise of deep learning technologies, numerous new segmentation methods have been proposed. The U-Net \cite{ronneberger2015unet}, based on the Encoder-Decoder architecture, innovatively utilizes skip connections to combine shallow and deep features, solving the problem of fine-grained information loss due to downsampling. On this basis, U-Net++\cite{zhou2018unet++} and ResUNet \cite{diakogiannis2020resunet} have been widely recognized for their enhanced performance in medical image segmentation. Recent studies have focused on enhancing model performance by improving various modules in the Encoder-Decoder architecture or introducing auxiliary supervision tasks.
CPFNet \cite{feng2020cpfnet} fuses global/multi-scale contextual information by combining a global pyramid guidance module and a scale-aware pyramid fusion module. PraNet \cite{fan2020pranet} progressively extends object regions by incorporating edge features through parallel partial decoders and reverse attention modules. TGANet \cite{tomar2022tganet} guides the model to learn additional feature representations by introducing supervision of textual labels. 
Swin-Unet \cite{cao2022swin-unet}, TransUNet \cite{chen2021transunet}, XBoundFormer \cite{wang2023xboundformer} enhance the performance by using various variants of Transformer Encoder. 

In medical images, the foreground (e.g. polyps, glands, lesions, etc.) and the background often do not have clear boundaries, but rather a ``soft boundary''. Furthermore, the lighting conditions and contrast in the image are often limited due to physical constraints and the complex reflective properties of the internal tissue. In recent years, many methods have been devoted to overcoming this challenge. 
SFA \cite{fang2019sfa} introduced an additional decoder for boundary prediction and employs a boundary-sensitive loss function to exploit the region-boundary relationship. 
BCNet \cite{yue2022bcnet} proposed a bilateral boundary extraction module that combines shallow contextual features, high-level positional features, and additional polyp boundary supervision to capture boundaries.
CFA-Net \cite{zhou2023cfa-net} designed a boundary prediction network to generate boundary-aware features, which are incorporated into the segmentation network using a layer-wise strategy.
MEGANet \cite{bui2024meganet} proposed an edge-guided attention module that uses the Laplacian operator to emphasize boundaries. 
Although these methods improve the model's focus on boundaries by explicitly introducing supervision related to boundaries, they do not fundamentally enhance the model's ability to spontaneously reduce the uncertainty of ambiguous regions. Therefore, in harsh environments, these methods exhibit weak robustness and offer limited performance improvement for the model. 
Moreover, the co-occurrence phenomenon is an easily overlooked challenge in medical image segmentation. Unlike objects that appear randomly in natural scenes, organs and tissues in medical images show a high degree of fixity and regularity, and thus co-occurrence widely exists \cite{chen2022co-occurrence}.
For example, in colonoscopy images, smaller polyps often appear in multiples simultaneously, and larger polyps are frequently accompanied by smaller polyps in their vicinity. This leads to the possibility that existing models may rely excessively on these contextual associations during the training process, rather than on the features of the polyps themselves. When a single polyp appears, models often have a tendency to predict multiple ones. This is precisely due to the failure of the model to learn to accurately distinguish between foreground and background, and between different entities in the image.


\section{Methodology}

\subsection{Network Architecture}
ConDSeg is a two-stage architecture. Figure \ref{fig:framework} shows its overall framework. In the first stage, we introduce the Consistency Reinforcement strategy to preliminarily train the Encoder, forcing the Encoder blocks to enhance robustness against adverse conditions such as weak illumination and low contrast, ensuring effective extraction of high-quality features in various scenarios. 
In the second stage, the learning rate for the Encoder is set at a lower level for fine-tuning. The whole ConDSeg network can be divided into four steps: 1) The ResNet-50 Encoder extracts feature maps \(f_1\) to \(f_4\) with varying semantic information at different levels. 2) The feature map \(f_4\), which carries deep semantic information, is fed into the Semantic Information Decoupling (SID) to be decoupled into feature maps enriched with foreground, background, and uncertainty region information, \(f_{fg}\), \(f_{bg}\), \(f_{uc}\). 3) Feature maps \(f_1\) to \(f_4\) are sent into the Contrast-Driven Feature Aggregation (CDFA) module, facilitating gradual fusion of multi-level feature maps guided by \(f_{fg}\) and \(f_{bg}\), and enhancing the representation of foreground and background features. 4) Decoder\(_s\), Decoder\(_m\) and Decoder\(_l\) each receive outputs from CDFA at specific levels to locate multiple entities within the image by size. The outputs from each decoder are fused to produce the final mask.

\subsection{Improving Boundary Distinction}
To overcome the challenge of blurred boundaries, the Consistency Reinforcement strategy and the Semantic Information Decoupling module are proposed, which will be elaborated in the following two subsections.

\subsubsection{Consistency Reinforcement}
The Encoder underpins the entire model's performance with its feature extraction capability and robustness. Therefore, in the first stage, our objective is to maximize the feature extraction capability of the Encoder and its robustness under weak illumination and low contrast scenarios.
To achieve this, we isolate the Encoder from the whole network and design a prediction head for it. We ensure that the structure of this prediction head is as simple as possible to avoid providing additional structures with feature-extraction capability beyond the Encoder blocks, thereby optimizing the Encoder performance as much as possible. We refer to this initial training network in the first stage as \(Net_0\).
For an input image \(X\), on the one hand, it is directly fed into \(Net_0\); on the other hand, \(X\) is strongly augmented to obtain \(X'\), which is then fed into \(Net_0\). These augmentation methods include randomly changing brightness, contrast, saturation, and hue, converting to grayscale images randomly and adding Gaussian blur. In this way, we simulate the variable environments in medical imaging \cite{7abdollahi2020data_aug,6goceri2023medical_data_aug}, such as weak illumination, low contrast and blur. This process can be represented as follows:
\begin{equation}
    \begin{aligned}
M_1 = Net_0(X), \quad M_2 = Net_0(\text{Aug}(X)).
\end{aligned}
\end{equation}
Here, \(\text{Aug}(\cdot)\) represents the strong augmentation operation. \(M_1\) and \(M_2\) respectively represent the predictions using the original image and the strongly augmented image. For \(M_1\) and \(M_2\), it is not only necessary for them to approximate the ground truth, but also need to maximize their similarity. This ensures the robustness of the Encoder in feature extraction, i.e., for the same content, the predicted mask should be consistent under any conditions and should not be affected by changes in lighting, contrast, etc.

Toward this, on the one hand, we jointly use Binary Cross-Entropy (BCE) Loss and Dice Loss to constrain \(M_1\) and \(M_2\):
\begin{equation}
    \mathcal{L}_{\text{BCE}}(Y, M) = -\frac{1}{N} \sum_{i=1}^N\left(Y_i \log \left(M_i\right) + \left(1-Y_i\right) \log \left(1-M_i\right)\right),
\end{equation}
\begin{equation}
    \mathcal{L}_{\text{Dice}}(Y, M) = 1 - 2 \times \frac{\sum_{i=1}^N M_i Y_i}{\sum_{i=1}^N M_i + \sum_{i=1}^N Y_i},
\end{equation}
\begin{equation}
    \mathcal{L}_{\text{mask}}(Y, M) = \mathcal{L}_{\text{BCE}}(Y, M) + \mathcal{L}_{\text{Dice}}(Y, M),
\end{equation}
where \(i\) represents the index of all pixels, \(M_i\) is the predicted value, and \(Y_i\) is the true label. \(N\) denotes the total number of pixels. Through the BCE Dice Loss, we consider both the pixel-level prediction accuracy and the issue of class imbalance in segmentation tasks. 
On the other hand, we design a novel consistency loss to maximize the similarity between \(M_1\) and \(M_2\) during the training process. 
When quantifying the consistency of model outputs, traditional methods such as KL divergence and JS divergence usually measure similarity from the perspective of probability distributions. These methods have a clear statistical significance in theory but may encounter numerical instability in practical applications, especially when predicted probabilities are close to 0 or 1. Logarithmic functions at these extreme values can cause numerical issues, which is particularly significant when dealing with large amounts of data. 
In contrast, our proposed consistency loss \(\mathcal{L}_{cons}\) is designed based on pixel-level classification accuracy, using simple binarization operation and Binary Cross-Entropy (BCE) loss calculation, directly comparing pixel-level differences between predicted masks \(M_1\) and \(M_2\). This makes the calculation simpler and avoids numerical instability, making it more robust for large-scale data. Specifically, we alternately binarize one of \(M_1\) and \(M_2\) as a reference and calculate the BCE Loss with the other. First, define a threshold \(t\) for the binarization operation, and the binarization function \(B(M, t)\) is defined as:
\begin{equation}
    B(M, t)= 
\begin{cases}
1 & \text{if } M \geq t, \\
0 & \text{otherwise}
\end{cases}
\end{equation}
The consistency loss \(\mathcal{L}_{cons}\) is given by the following formula, which calculates the average of the BCE Loss of \(M_2\) with \(M_1\) as the reference, and the BCE Loss of \(M_1\) with \(M_2\) as the reference:
\begin{align}
    \mathcal{L}_{cons}\left(M_1, M_2\right) ={}& \frac{1}{2}\bigg(\mathcal{L}_{BCE}\left(B\left(M_2, t\right), M_1\right) \nonumber \\
    &+ \mathcal{L}_{BCE}\left(B\left(M_1, t\right), M_2\right)\bigg).
\end{align}
Finally, the overall loss for the first stage is as follows:
\begin{equation}
\mathcal{L}_{stage_1}=\mathcal{L}_{mask_1}+\mathcal{L}_{mask_2}+\mathcal{L}_{cons}.
\end{equation}

\subsubsection{Semantic Imformation Decoupling}
The Semantic Information Decoupling (SID) module aims to decouple high-level feature map outputted by the Encoder into three feature maps enriched with information on the foreground, background, and regions of uncertainty, respectively. By constraining with loss functions, the SID is encouraged to learn to reduce uncertainty, allowing for precise differentiation between foreground and background.

SID begins with three parallel branches, each consisting of multiple CBR blocks. The feature map \(f_4\) from the Encoder is inputted into these three branches to obtain three feature maps with different semantic information, \(f_{fg}, f_{bg}, f_{uc}\), which are enriched with the foreground, background, and uncertainty region features respectively. The three feature maps are then predicted by an auxiliary head to produce masks for the foreground (\(M^{fg}\)), background (\(M^{bg}\)), and regions of uncertainty (\(M^{fg}\)). Ideally, the value at each pixel of these three masks should be a definitive 0 or 1, indicating that the category to which the pixel belongs is certain. Moreover, the three feature maps should exhibit a complementary relationship, i.e., for any pixel index \(i\), the three masks \(M^{fg}\), \(M^{bg}\), and \(M^{uc}\) should satisfy:
\begin{equation}
    \left\{\begin{aligned}
&M_i^{fg}, M_i^{bg}, M_i^{uc} \in \{0,1\} \\
&M_i^{fg} + M_i^{bg} + M_i^{uc} = 1
\end{aligned}\right.
\end{equation}
Therefore, our optimization goal is for \(M^{fg}\) and \(M^{bg}\) to approximate the ground truth \(Y\) and its negation \((1-Y)\), respectively. Meanwhile, \(M^{fg}\), \(M^{bg}\), and \(M^{uc}\) should approximately satisfy the aforementioned complementary.

On the one hand, we optimize \(M^{fg}\) and \(M^{bg}\) towards the true labels using the BCE Dice Loss, denoted as \(\mathcal{L}_{fg}\) and \(\mathcal{L}_{bg}\). Additionally, considering that small-sized entities occupy very few pixels and thus their prediction accuracy has a minor impact on the overall loss, we designed a pair of dynamic penalty terms, \(\beta_1, \beta_2\):
\begin{equation}
    \beta_1 = \frac{1}{\tanh\left(\frac{\sum_{i=1}^N M_i^{fg}}{N}\right)}, \quad \beta_2 = \frac{1}{\tanh\left(\frac{\sum_{i=1}^N M_i^{bg}}{N}\right)},
\end{equation}
where \(N\) represents the total number of pixels and \(i\) is the index for all pixels, ranging from 1 to \(N\). Essentially, \(\beta_1\) and \(\beta_2\) are penalty terms calculated based on the area ratio of the prediction results, designed to give more attention to entities with smaller areas in the loss function, thus improving the stability of the loss function. Therefore, with the addition of the penalty terms, \(\mathcal{L}_{fg}\) and \(\mathcal{L}_{bg}\) are transformed into \(\beta_1 \mathcal{L}_{fg}\) and \(\beta_2 \mathcal{L}_{bg}\).

On the other hand, considering that the outputs \(M^{fg}\), \(M^{bg}\), and \(M^{uc}\) from the auxiliary head are in fact probability distributions, we designed a simple yet effective loss function to quantify their degree of complementarity:
\begin{equation}
    \mathcal{L}_{compl} = \frac{1}{N} \sum_{i=1}^N\left(M_i^{fg} \cdot M_i^{bg} + M_i^{fg} \cdot M_i^{uc} + M_i^{bg} \cdot M_i^{uc}\right).
\end{equation}
Here, \(M_i^{fg}\), \(M_i^{bg}\), and \(M_i^{uc}\) represent the probability of the \(i\)-th pixel being predicted to be foreground, background, and an uncertainty region, respectively. This formulation provides a direct and effective constraint on the SID's predictions. By normalizing the loss by the total number of pixels, the scale of the loss function is made independent of the input image size, keeping its value consistently within the range \([0,1]\), thus fully ensuring the stability of training.

Overall, \(\mathcal{L}_{compl}\) enforces the complementarity of the prediction results from the three branches, ensuring each pixel is categorized as belonging exclusively to one of the foreground, background, or uncertainty regions. Additionally, \(\beta_1 \mathcal{L}_{fg}\) and \(\beta_2 \mathcal{L}_{bg}\) are used to constrain the accuracy of the foreground and background branches.  With these constraints during training, the model gradually reduces the uncertainty regions, and enhances the prominence of the foreground and background within the feature maps \(f_{fg}\) and \(f_{bg}\).

\begin{table*}[t]
\small
\centering
\renewcommand\arraystretch{0.9} 
\caption{Comparison with other methods on the Kvasir-Sessile, Kvasir-SEG and GlaS datasets.}
\label{tab:comp_3}
\begin{tabular}{lcccccccccccc}
\toprule
\multirow{2}{*}{Methods}& \multicolumn{4}{c}{Kvasir-Sessile} & \multicolumn{4}{c}{Kvasir-SEG} & \multicolumn{4}{c}{GlaS} \\
\cmidrule(r){2-5} \cmidrule(r){6-9} \cmidrule(r){10-13}
 & mIoU & mDSC & Rec. & Prec. & mIoU & mDSC & Rec. & Prec. & mIoU & mDSC & Rec. & Prec.\\
\midrule
U-Net & 23.1 & 33.8 & 45.1 & 46.6 & 65.5 & 75.8 & 83.6 & 77.6 & 75.8 & 85.5 & 90.3 & 82.8 \\
U-Net++ & 38.4 & 50.2 & 62.5 & 51.8 & 67.9 & 77.2 & 86.5 & 77.6 & 77.6 & 86.9 & 89.6 & 85.5 \\
Attn U-Net & 27.5 & 38.9 & 59.0 & 44.4 & 67.6 & 77.4 & 83.9 & 79.9 & 76.6 & 85.9 & 91.8 & 82.2 \\
PraNet & 66.7 & 77.4 & 80.7 & 82.4 & 83.0 & 89.4 & 90.6 & 91.3& 71.8& 83.0& 90.9& 78.0\\
TGANet & 74.4& 82.0& 79.3 & \underline{85.9}& 83.3 & 89.8& 91.3& 91.2 & 71.8 & 84.7 & 86.9 & 80.2 \\
DCSAU-Net & 72.6 & 81.1 & 65.6 & 62.9 & 83.5 & 88.9 & 89.5 & 89.5 & 77.6 & 86.5 & \underline{93.0}& 82.5 \\
XBFormer & 73.6 & 81.1 & \underline{87.2}& 76.3 & 83.8& 88.9 & 89.8 & 87.2 & 73.7 & 84.3 & 84.0 & 85.7 \\
CASF-Net & 60.5 & 72.4 & 78.0 & 74.8 & 81.7 & 88.7 & 89.2 & 88.2 & 78.4& 87.2& 91.3 & 85.9\\
 DTAN& \underline{76.4}& \underline{84.2}& 84.2& \underline{85.9}& \underline{84.1}& \underline{90.4}& \underline{91.6}& \textbf{92.0}& \underline{78.5}& \underline{87.9}& 85.8& \underline{90.2}\\
\midrule
\textbf{Ours} & \textbf{81.2}& \textbf{89.1} & \textbf{90.1} & \textbf{90.0} & \textbf{84.6} & \textbf{90.5} & \textbf{92.3} & \underline{91.7}& \textbf{85.1} & \textbf{91.6} & \textbf{93.5} & \textbf{90.5}\\
\bottomrule
\end{tabular}
\end{table*}

\subsection{Overcoming Co-occurrence}
\begin{figure}[t]
    \centering
    \includegraphics[width=0.75\linewidth]{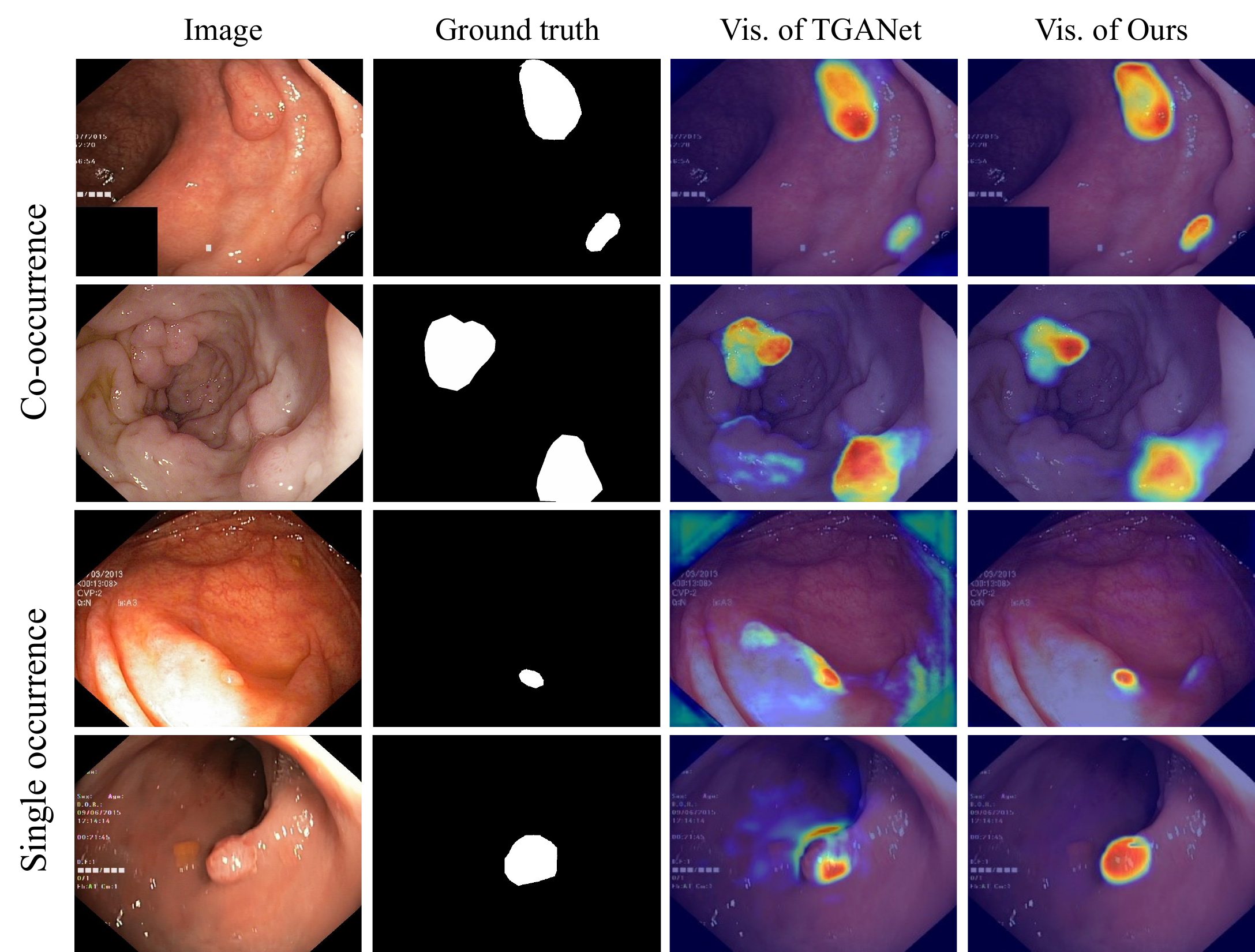}
    \caption{The Grad-CAM visualization of different methods. It illustrates the attention areas of our model and TGANet in cases of co-occurrence and individual appearances.}
    \label{fig:cam_vis}
\end{figure}
Co-occurrence phenomena are widespread in medical imaging, which often cause models to learn incorrect patterns, thus degrading their performance. Figure \ref{fig:cam_vis} shows a visual comparison between TGANet \cite{tomar2022tganet} and our method on several examples from the Kvasir-SEG dataset \cite{jha2020kvasir-seg}. For both, we utilize Grad-CAM \cite{selvaraju2017grad-cam} to visualize convolutional layers near the output layer, intuitively demonstrating the focus of the models. For co-occurring polyps, both models can achieve accurate localization. However, when the polyp appears alone, TGANet tends to mistakenly predict multiple polyps. In contrast, our method avoids the misleading effects of co-occurrence phenomena. It remains capable of accurately recognizing and segmenting when a single polyp appears. In our framework, we overcome co-occurrence phenomena through the specially designed Contrast-Driven Feature Aggregation module and Size-Aware Decoder, which will be detailed in the following two subsections.

\subsubsection{Contrast-Driven Feature Aggregation}

The Contrast-Driven Feature Aggregation (CDFA) module aims to utilize contrastive features of the foreground and background decoupled by SID to guide multi-level feature fusion. Moreover, it helps the model better distinguish between entities to be segmented and the complex background environments. Specifically, \(f_1\) to \(f_4\) outputted from the Encoder blocks are first pre-enhanced through a set of dilated convolutional layers to obtain \(e_1\) to \(e_4\). Subsequently, CDFA achieve multi-level feature fusion guided by contrastive features (\(f_{fg}\), \(f_{bg}\)).


\begin{figure}[t]
    \centering
    \includegraphics[width=0.9\linewidth]{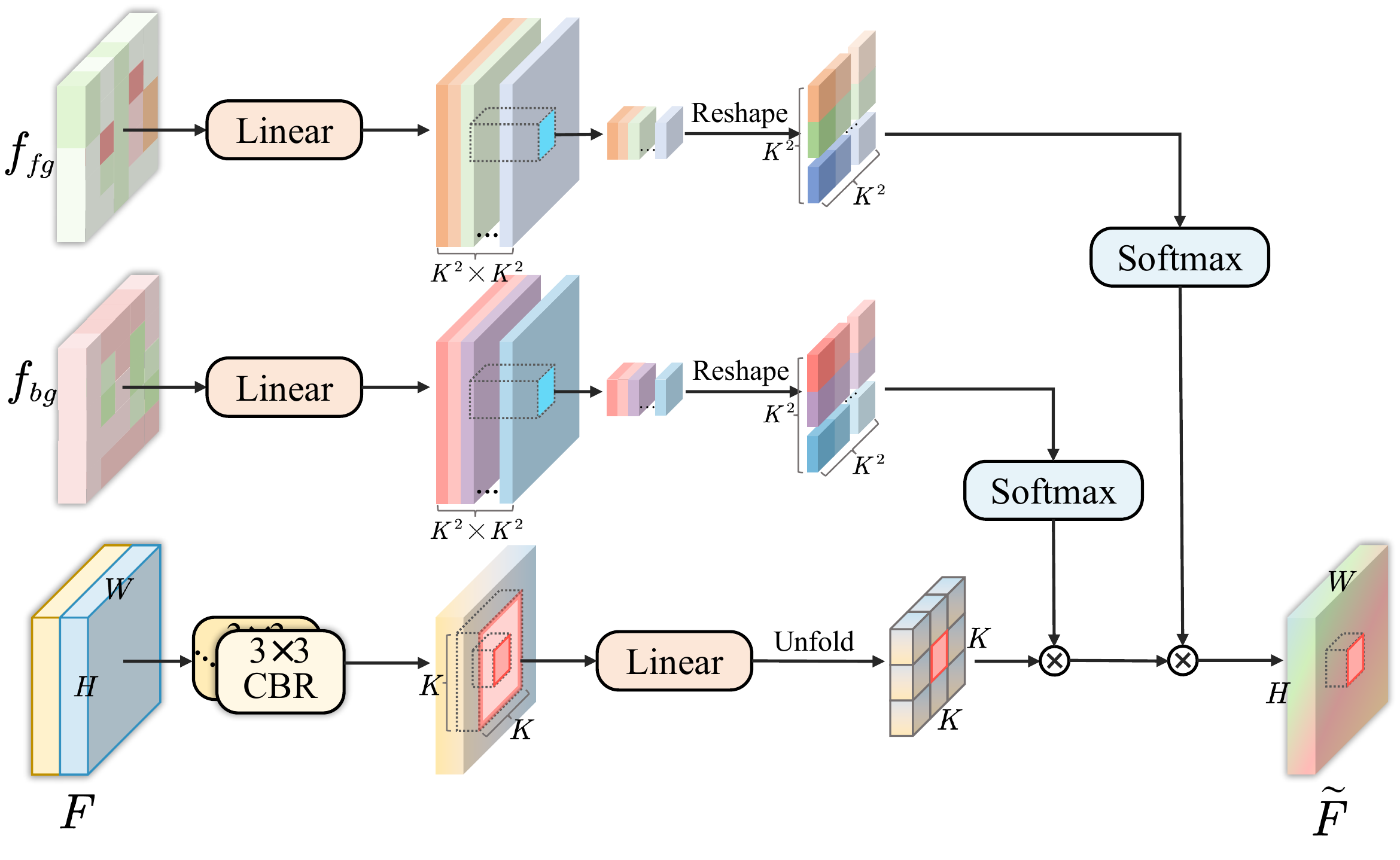}
    \caption{The structure of CDFA.}
    \label{fig:cdfa}
    \vspace{-0.3cm}
\end{figure}


On the one hand, the output from the previous level and the lateral feature map (i.e. \(e_i, i=1,2,3,4\)) are concatenated along the channel direction to form \(F\) . Then \(F\) is fed into CDFA (except \(e_4\), which is directly inputted into CDFA). On the other hand, \(f_{fg}\) and \(f_{bg}\) are adjusted to match the dimensions of \(F\) through convolutional layers and bilinear upsampling before being fed into CDFA.

Figure \ref{fig:cdfa} shows the structure of CDFA. \(f_{fg}\) and \(f_{bg}\) are both derived from the decoupling of deep features, where the features at each spatial position are representative enough to generate attention weights for locally aggregating neighboring features. At each spatial location \((i, j)\), CDFA calculates the attention weights within a \(K \times K\) window centered at \((i, j)\), by incorporating details from the foreground and background. Specifically, given the input feature map \(F \in \mathbb{R}^{H \times W \times C}\), it is first subjected to preliminary fusion through multiple CBR blocks. Subsequently, this \(C\)-dimensional feature vector is mapped to a value vector \(V \in \mathbb{R}^{H \times W \times C}\) through a linear layer weight \(W_V \in \mathbb{R}^{C \times C}\). The value vector \(V\) is then unfolded over each local window, preparing to aggregate neighborhood information for each position. Let \(V_{\Delta_{i, j}} \in \mathbb{R}^{C \times K^2}\) represent all the values within the local window centered at \((i, j)\), defined as:
\begin{equation}
    V_{\Delta_{i, j}}=\left\{V_{i+p-\left\lfloor\frac{K}{2}\right\rfloor, j+q-\left\lfloor\frac{K}{2}\right\rfloor}\right\}, 0 \leq p, q<K.
\end{equation}
The foreground and background feature maps (\(f_{fg}\) and \(f_{bg}\)) are then processed through two distinct linear layers to generate corresponding attention weights \(A_{fg} \in \mathbb{R}^{H \times W \times K^2}\) and \(A_{bg} \in \mathbb{R}^{H \times W \times K^2}\). The calculation of attention weights can be represented as:
\begin{equation}
    A_{fg} = W_{fg} \cdot f_{fg}, \quad A_{bg} = W_{bg} \cdot f_{bg},
\end{equation}
where \(W_{fg} \in \mathbb{R}^{C \times K^4}\) and \(W_{bg} \in \mathbb{R}^{C \times K^4}\) are the linear transformation weight matrices for the foreground and background feature maps, respectively. Subsequently, the foreground and background attention weights at position \((i, j)\) are reshaped into \(\hat{A}_{fg_{i, j}} \in \mathbb{R}^{K^2 \times K^2}\) and \(\hat{A}_{bg_{i, j}} \in \mathbb{R}^{K^2 \times K^2}\), and both of them are activated by a Softmax function. Then, the unfolded value vector \(V_{\Delta_{i, j}}\) is weighted in two steps:
\begin{equation}
    \tilde{V}_{\Delta_{i, j}} = \text{Softmax}( \hat{A}_{fg_{i, j}}) \otimes (\text{Softmax}( \hat{A}_{bg_{i, j}}) \otimes V_{\Delta_{i, j}}).
\end{equation}
Here, \(\otimes\) denotes matrix multiplication. Finally, the weighted value representations are densely aggregated to obtain the final output feature map. Specifically, the aggregated feature at position \((i, j)\) is:
\begin{equation}
    \overset{\sim}{F}_{i,j} = \sum_{0 \leq m,n < K} \overset{\sim}{V}_{\Delta_{i + m - \lfloor\frac{K}{2}\rfloor,j + n - \lfloor\frac{K}{2}\rfloor}}^{i,j}.
\end{equation}

\subsubsection{Size-Aware Decoder}
We designed a simple but effective Size-Aware Decoder (SA-Decoder), whose structure is detailed in the \textit{Supplementary Material}.
SA-Decoder implements separate predictions by distributing entities of different sizes in different layers.
Among the features outputted by multiple CDFAs, the shallow-level feature maps contain more fine-grained information, making them suitable for predicting smaller-sized entities. As the layers deepen, the feature maps incorporate increasing amounts of global information and higher-level semantics, making them more suitable for predicting larger-sized entities.

Therefore, we establish three Decoders for small, medium, and large sizes: Decoder\(_s\), Decoder\(_m\), and Decoder\(_l\), each receives features from the two adjacent CDFAs, namely \(\widetilde{F}_1\) and \(\widetilde{F}_2\), \(\widetilde{F}_2\) and \(\widetilde{F}_3\), \(\widetilde{F}_3\) and \(\widetilde{F}_4\), respectively. 

The outputs of the three Decoders are then concatenated and fused along the channel dimension. Then the predicted mask is produced through a Sigmoid function. Through the collaborative work of multiple parallel SA-Decoders, ConDSeg is able to accurately distinguish individuals of different sizes. The model is capable of both precisely segmenting large entities and accurately locating small entities.

\subsection{Overall Optimization and Training Process}

ConDSeg's training is a two-stage process. The first stage focuses on enhancing the feature extraction capability of the Encoder and its robustness to adverse conditions. In the second stage, we set the learning rate of Encoder at a lower level and optimize entire model. Formally, \(\mathcal{L}_{stage_2}\) is as follows:
\begin{equation}
    \mathcal{L}_{stage_2} = \mathcal{L}_{mask} + \beta_1 \mathcal{L}_{fg} + \beta_2 \mathcal{L}_{bg} + \mathcal{L}_{compl}.
\end{equation}


\begin{table}[t]
\small
\centering
\renewcommand\arraystretch{0.9} 
\caption{Comparison with other methods on the ISIC-2016 and ISIC-2017 datasets.}
\label{tab:comp_2}
\begin{tabular}{lcccc}
\toprule
\multirow{2}{*}{Method} & \multicolumn{2}{c}{ISIC-2016} & \multicolumn{2}{c}{ISIC-2017} \\
\cmidrule(lr){2-3} \cmidrule(lr){4-5}
 & mIoU & mDSC & mIoU & mDSC \\
\midrule
U-Net & 83.6 & 90.3 & 73.7 & 82.8 \\
CENet& 84.6& 90.9& 76.4&84.8\\
CPFNet & 84.2 & 90.7 & 76.2 & 84.7 \\
FAT-Net & 85.3 & 91.6 & 76.5 & 85.0 \\
DCSAU-Net & 85.3 & 91.4 & 76.1 & 85.0 \\
EIU-Net& \underline{85.5}& \underline{91.9}& \underline{77.1}& \underline{85.5}\\
\midrule
Ours & \textbf{86.8} & \textbf{92.5} & \textbf{80.9} & \textbf{88.3} \\
\bottomrule
\end{tabular}
\end{table}

\section{Experiments}
\subsection{Experimental Setup}

To verify the performance and general applicability of our ConDSeg in the field of medical image segmentation, we conducted experiments on five challenging public datasets: Kvasir-SEG \cite{jha2020kvasir-seg}, Kvasir-Sessile \cite{jha2021sessile}, GlaS \cite{sirinukunwattana2017glas}, ISIC-2016 \cite{gutman2016isic-2016}, and ISIC-2017 \cite{codella2018isic-2017}, covering subdivision tasks across three medical image modalities. Detailed information about the datasets is shown in the \textit{Supplementary Material}.

All experiments were conducted on an NVIDIA GeForce RTX 4090 GPU, with the image size adjusted to \(256 \times 256\) pixels. Simple data augmentation strategies, including random rotation, vertical flipping and horizontal flipping were used. The batch size was set to 4, and the Adam optimizer \cite{kingma2014adam} was used for optimization. We use the ResNet-50 \cite{he2016resnet} as the default encoder, which can also be replaced with other backbones such as Transformers (see the \textit{Supplementary Material}). In the first stage, the learning rate is set to 1\(e\)-4. In the second stage, we load the weights of the Encoder and set its learning rate to a lower 1\(e\)-5, while for the rest of the network, the learning rate is set to 1\(e\)-4. The window size for CDFA is set to 3. We use standard medical image segmentation metrics such as mean Intersection over Union (mIoU), mean Sørensen-Dice coefficient (mDSC), Recall, and Precision.

\subsection{Comparison with Other State-of-the-Art}

\begin{figure}[h]
    \centering
    \includegraphics[width=1\linewidth]{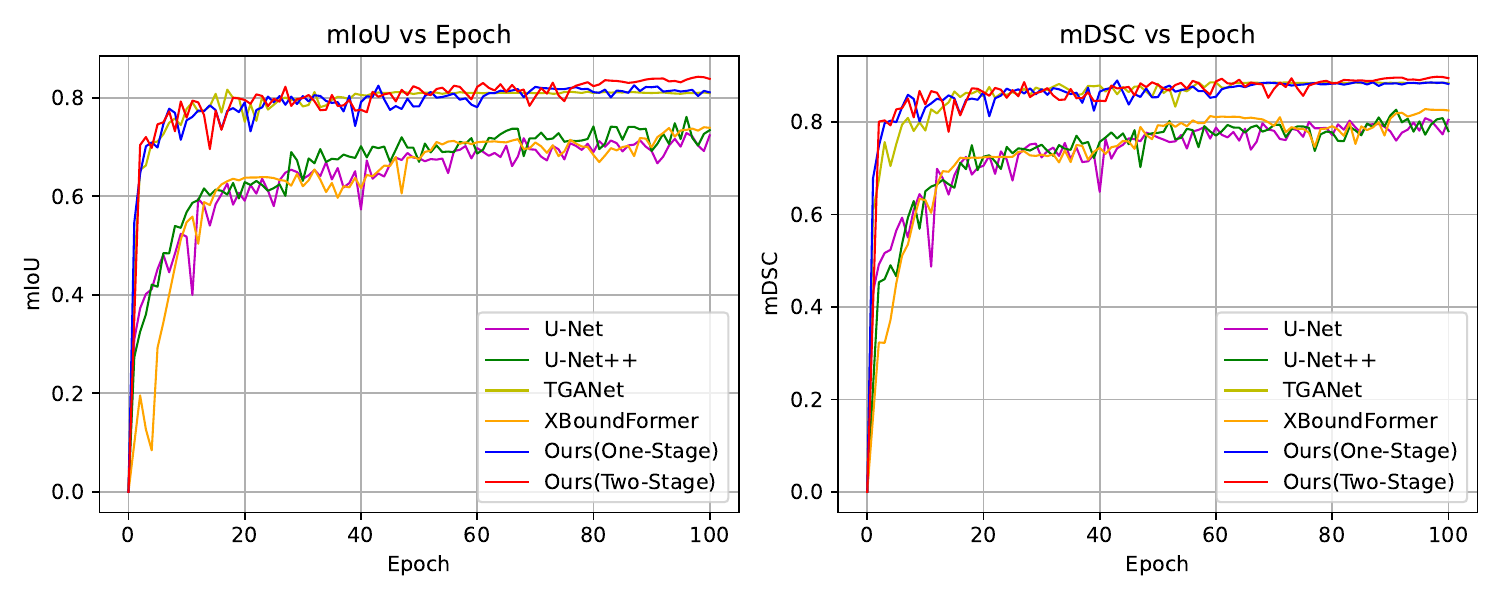}
    \caption{Comparison of convergence curves with other methods in the training process on the Kvasir-SEG dataset.}
    \label{fig:convergence}
\end{figure}

Tables \ref{tab:comp_3} and \ref{tab:comp_2} detail the comparison of our method with the following state-of-the-art methods: U-Net\cite{ronneberger2015unet},U-Net++\cite{zhou2018unet++}, Attn U-Net \cite{oktay2018attention-unet}, CENet\cite{gu2019cenet}, CPFNet\cite{feng2020cpfnet}, PraNet\cite{fan2020pranet}, FAT-Net\cite{wu2022fat-net}, TGANet\cite{tomar2022tganet}, DCSAU-Net\cite{xu2023dcsau-net}, XBoundFormer\cite{wang2023xboundformer}, CASF-Net\cite{zheng2023casf-net}, EIU-Net\cite{yu2023eiu-net}, DTAN \cite{zhao2024dtan}. On all five datasets, our method achieves the best segmentation performance.
Additionally, we compare the convergence speed during the training in Figure \ref{fig:convergence}. Here, One-Stage and Two-Stage respectively represent ConDSeg trained from scratch and trained with the pre-trained weights (using CR) loaded. The figure shows that ConDSeg can reach an advanced level even with only one stage training. When using the complete ConDSeg framework (Two-Stage), our method achieves both the fastest convergence speed and best performance.


\subsection{Ablation Study}
To verify the effectiveness of our proposed Consistency Reinforcement training strategy and all the proposed modules, we conducted comprehensive ablation experiments, which will be elaborated in the following two subsections.

\subsubsection{Training Strategy}

\begin{table}[htbp]
\small
\centering
\renewcommand\arraystretch{0.9} 
\caption{Ablation experiments on training strategies.}
\label{tab:ablation_train}
\begin{tabular}{ccccccc}
\toprule
 \multicolumn{2}{c}{Stage 1}& \multirow{2}{*}{Stage 2} & \multicolumn{4}{c}{Metric} \\
\cmidrule{1-2}  \cmidrule{4-7} 
\(Net_0\) & CR &  & mIoU & mDSC & Rec. & Prec. \\
\midrule
\checkmark &  &  & 77.3 & 85.1 & 86.7 & 88.6 \\
\checkmark & \checkmark &  & 78.4 & 86.1 & 87.1 & 89.8 \\
 &  & \checkmark & 82.7 & 88.9 & 91.8 & 90.2 \\
\checkmark &  & \checkmark & 82.9 & 89.2 & 92.0 & 90.2 \\
\checkmark & \checkmark & \checkmark & \textbf{84.6} & \textbf{90.5} & \textbf{92.3} & \textbf{91.7} \\
\bottomrule
\end{tabular}
\end{table}

On the Kvasir-SEG dataset, we designed 5 sets of ablation experiments for ConDSeg's training strategy:
1) Direct training of \(Net_0\) (proposed in 3.2.1) and using \(Net_0\) for prediction.
2) Training \(Net_0\) with the CR strategy and using \(Net_0\) for prediction.
3) Directly training the entire ConDSeg network, without using a two-stage strategy.
4) Employing the two-stage strategy. In the first stage, \(Net_0\) is used to preliminarily train the Encoder, and in the second stage, the entire network is trained. However, CR is not used in the first stage.
5) Using the two-stage strategy, with CR employed in the first stage.
Among these 5 experiments, 1) and 2) are aimed at comparing and verifying the effectiveness of CR in enhancing the robustness of the Encoder. 3) to 5) are designed to verify the effectiveness of using a two-stage strategy and the additional benefits of employing the proposed CR on network performance.
As shown in Table \ref{tab:ablation_train}, the experimental results under settings 1) to 5) are sequentially presented. 
It indicates that: the CR strategy significantly improves the Encoder's feature extraction capability and its robustness to adverse conditions; adopting a two-stage training approach and utilizing the CR strategy in the first stage significantly enhances model performance, achieving the best results.

\subsubsection{Proposed Modules}

\begin{table}[htbp]
\small
\centering
\renewcommand\arraystretch{0.9} 
\caption{Ablation experiments on different modules.}
\label{tab:abaltion_modules}
\begin{tabular}{ccccc}
\toprule
\multicolumn{3}{c}{Modules} & \multicolumn{2}{c}{Metrics} \\
\cmidrule(lr){1-3} \cmidrule(lr){4-5}
Baseline & SID \& CDFA & SA-Decoder & mIoU & mDSC  \\
\midrule
\checkmark &   &  & 78.3 & 87.5 \\
\checkmark & \checkmark &  & 83.3 & 90.4 \\
\checkmark & \checkmark  & \checkmark & \textbf{85.1} & \textbf{91.6}  \\
\bottomrule
\end{tabular}
\end{table}

To validate the effectiveness of the proposed modules on model performance, we set up 3 groups of ablation experiments on the GlaS dataset. All experiments were conducted on the basis of the Encoder already enhanced by the CR strategy in the first stage. The baseline refers to that in ConDSeg, the SID is removed and the CDFA is replaced with normal convolutional layers, and a single-scale Decoder is used. Its specific structure is detailed in the \textit{Supplementary Material}. 
Our 4 experiment setups are: 1) Baseline.  2) Adding the SID and CDFA modules. 3) Further adopting the SA-Decoder, i.e., the complete ConDSeg architecture. According to Table \ref{tab:abaltion_modules}, both mIoU and mDSC are significantly improved when modules are added incrementally and are optimal when all modules are used.

More visualization and results of the ablation experiments are detailed in the \textit{Supplementary Material}.
\section{Conclusion}
We propose a general medical image segmentation framework, ConDSeg, which alleviates the challenges posed by soft boundaries and co-occurrence phenomena widespread in medical imaging. For issues related to ambiguous boundaries, as well as poor illumination and low contrast, our preliminary training strategy, Consistency Reinforcement, is specifically aimed at enhancing the Encoder's robustness to adverse conditions. Furthermore, we design the Semantic Information Decoupling module to decouple features from the Encoder into three parts, foreground, background and regions of uncertainty, gradually acquiring the ability to reduce uncertainty during training. 
To address the challenge of co-occurrence, the Contrast-Driven Feature Aggregation module uses foreground and background information to guide the fusion of multi-level feature maps and enhance the key features, facilitating the model further distinguishing between foreground and background. 
Moreover, we introduce the Size-Aware Decoder to achieve precise localization of multiple entities of different sizes within images. Across five widely used medical image segmentation datasets of various modalities, our ConDSeg achieves state-of-the-art performance, verifying the advanced nature and universality of our framework.

\bibliography{aaai25}

\newpage
\section{Supplementary Material}
\subsection{Dataset}
To verify the performance and general applicability of our ConDSeg in the field of medical image segmentation, we conducted experiments on five challenging public datasets: Kvasir-SEG \cite{jha2020kvasir-seg}, Kvasir-Sessile \cite{jha2021sessile}, GlaS \cite{sirinukunwattana2017glas}, ISIC-2016 \cite{gutman2016isic-2016}, and ISIC-2017 \cite{codella2018isic-2017}, covering subdivision tasks in three modalities of medical image. Detailed information on the datasets is shown in Table \ref{tab:dataset}. 
For Kvasir-SEG, we followed the official recommendation, using a split of 880/120 for training and validation. Kvasir-Sessile, a challenging subset of Kvasir-SEG, adopted the widely used split of 156/20/20 for training, validation, and testing as in \cite{tomar2022tganet}, \cite{dong2024tgediff}, etc. For GlaS, we used the official split of 85/80 for training and validation. For ISIC-2016, we utilized the official split of 900/379 for training and validation. For ISIC-2017, we also followed the official recommendation, using a split of 2000/150/600 for training, validation and testing.

\begin{table*}[h]
\centering
\renewcommand{\arraystretch}{0.8}
\caption{Detailed information of the five datasets.}
\label{tab:dataset}
\begin{tabular}{lcccc}
\toprule
\thead{Dataset} & \thead{Modality} & \thead{Anatomic Region} & \thead{Segmentation Target} & \thead{Data Volume} \\
\midrule
Kvasir-SEG & endoscope & colon & polyp & 1000 \\
\addlinespace[3pt]
Kvasir-Sessile & endoscope & colon & polyp & 196 \\
\addlinespace[3pt]
GlaS & whole-slide image (WSI) & colorectum & gland & 165 \\
\addlinespace[3pt]
ISIC-2016 & dermoscope & skin & malignant skin lesion & 1279 \\
\addlinespace[3pt]
ISIC-2017 & dermoscope & skin & malignant skin lesion & 2750 \\
\bottomrule
\end{tabular}

\end{table*}

\subsection{Supplementary Ablation Experiments}

During training, the threshold \(t\) is used to binarize \(M_1\) and \(M_2\). Similarly, during inference stage, the final output of the network also needs to be binarized to obtain segmentation results. If the value of \(t\) is too high, it will lead to over-segmentation and more false positives, while if the value of \(t\) is too low, it will lead to under-segmentation and false negatives. A common choice is to set \(t\) to 0.5 . In this regard, we trained and validated the model on the ISIC-2016 dataset with different values of \(t\). As shown in Figure 1, the model performs best when \(t\) is around 0.5 .

\begin{figure}[h]
    \centering
    \includegraphics[width=1\linewidth]{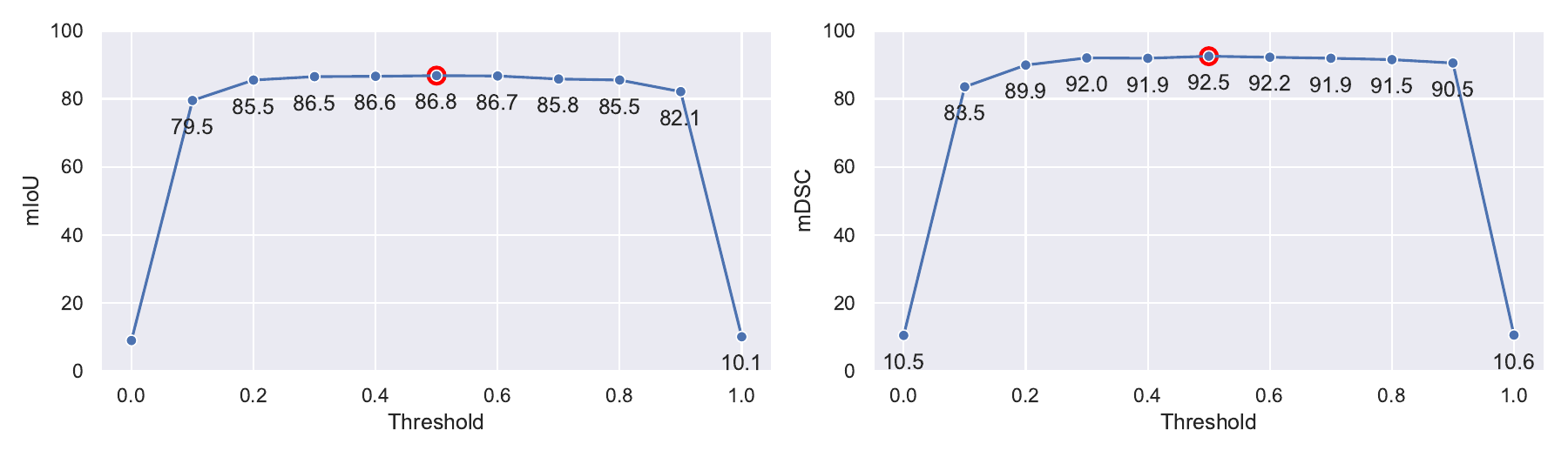}
    \caption{Model performance on ISIC-2016 dataset when different
thresholds (\(t\)) are selected}
    \label{fig:enter-label}
\end{figure}

\begin{figure}[H]
    \centering
    \includegraphics[width=1\linewidth]{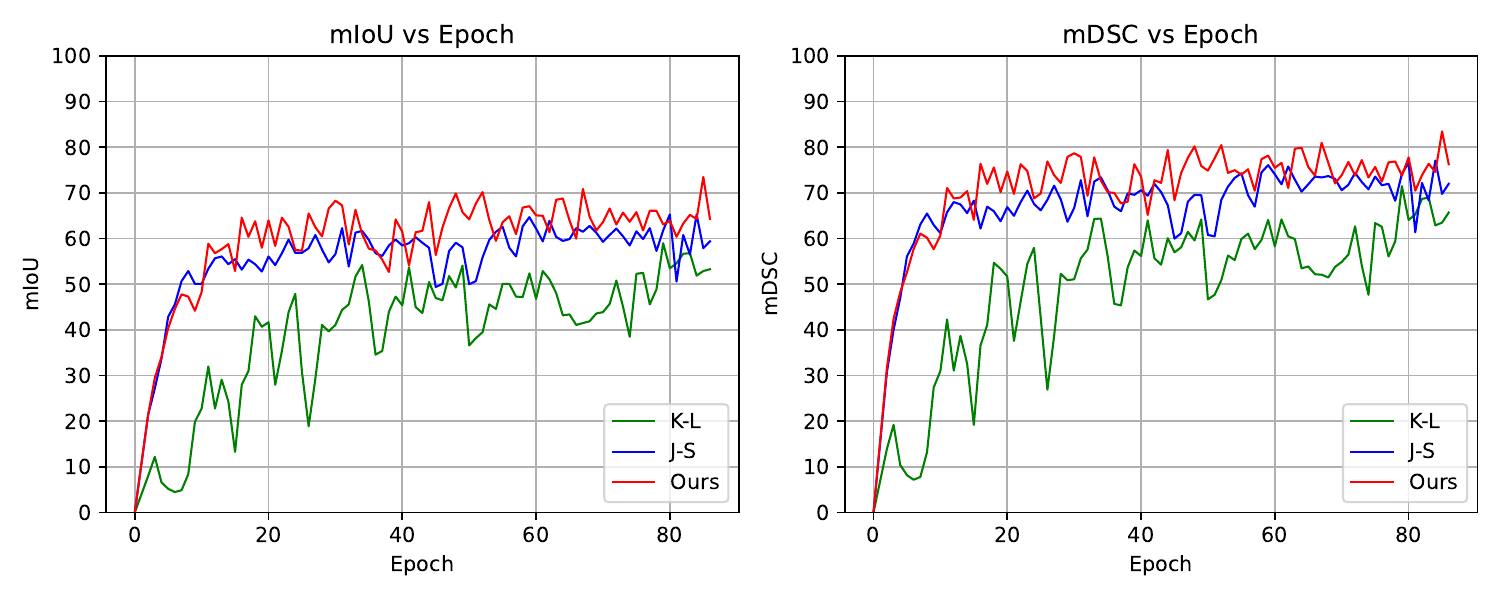}
    \caption{Comparison of convergence curves when training \(Net_0\) with different loss functions to optimise the consistency of the predictions. Here, K-L and J-S stand for K-L divergence and J-S divergence, respectively. }
    \label{fig:loss}
\end{figure}

To validate the superiority of our proposed \(\mathcal{L}_{cons}\), we conducted a set of experiments on the Kvasir-Sessile dataset, comparing it with Kullback-Leibler (K-L) divergence and Jensen-Shannon (J-S) divergence as the loss functions. Figure \ref{fig:loss} shows the convergence curves of \(Net_0\) when trained using different loss functions. Clearly, when using our \(\mathcal{L}_{cons}\), the convergence curve is the most stable and exhibits much better performance.

To explore the impact of the number of training epochs in stage 1 on the final performance of ConDSeg, we conducted five sets of experiments on the Kvasir-SEG dataset. We set the number of training epochs in stage 1 to 5, 10, 20, 50, and 100, respectively. As shown in Table \ref{tab:training_performance}, initially, as the number of training epochs in stage 1 increases, the performance of ConDSeg noticeably improves. However, from 50 to 100 epochs, the network performance essentially reached saturation, with no significant changes. This further proves the effectiveness of our proposed two-stage training strategy.

\begin{table}[H]
\centering
\small
\caption{Model performance across different training epochs in stage 1. Here ep. stands for the number of epochs.}
\label{tab:training_performance}
\begin{tabular}
{lccccc}
\toprule
Stage 1 & 5 ep. & 10 ep. & 20 ep. & 50 ep. & 100 ep. \\
\midrule
mIoU & 82.9 & 83.5 & 84.0 & 84.4 & 84.6 \\
mDSC & 89.0 & 89.4 & 90.1 & 90.6 & 90.5 \\
\bottomrule
\end{tabular}
\end{table}

To explore the impact of different window sizes (\(K\)) in the CDFA on model performance, we conducted four sets of experiments on the Kvasir-Sessile dataset, with window sizes set to 1, 3, 5, and 7, respectively. As shown in Table \ref{tab:window_size}, our model achieved good performance under all settings. However, a larger window size does not necessarily lead to a more significant improvement in model performance. The best performance was observed when \(K=3\).

\begin{table}[H]
\centering
\small
\caption{Impact on model performance when setting different window sizes (\(K\)) for CDFA on the Kvasir-Sessile dataset. The best results are achieved when \(K\)=3.}
\label{tab:window_size}
\begin{tabular}{ccccc}
\toprule
Window Size & mIoU & mDSC & Recall & Precision \\
\midrule
\(K\)=1 & 80.1 & 88.4 & 89.4 & 88.8 \\
\(K\)=3 & 81.2 & 89.1 & 90.1 & 90.0 \\
\(K\)=5 & 77.6 & 86.0 & 92.1 & 84.3 \\
\(K\)=7 & 77.1 & 85.9 & 91.5 & 85.5 \\
\bottomrule
\end{tabular}
\end{table}

\begin{figure}[H]
    \centering
    \includegraphics[width=1\linewidth]{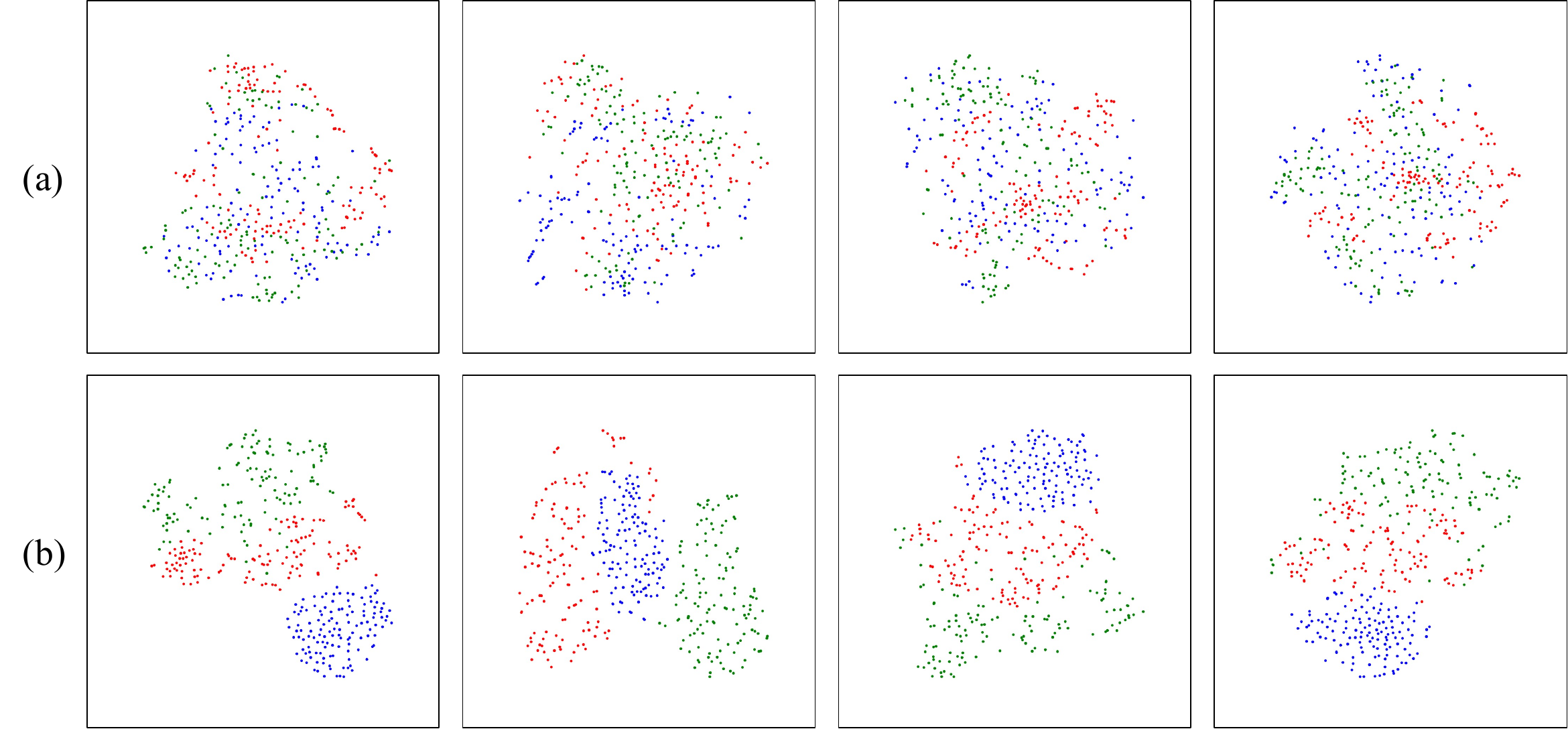}
    \caption{The t-SNE Visualisation of \(f_{fg}\), \(f_{bg}\) and \(f_{uc}\). (a) represents the results for training ConDSeg without the auxiliary head as well \(\mathcal{L}_{fg}\), \(\mathcal{L}_{bg}\) and \(\mathcal{L}_{compl}\), while (b) represents the results when they are used. Each column is the result of a different image. Here, green represents \(f_{fg}\),blue represents \(f_{bg}\), and red represents \(f_{uc}\).}
    \label{fig:tsne}
\end{figure}

To illustrate the effectiveness of the method for optimizing the Semantic Information Decoupling (SID) module, Figure \ref{fig:tsne} shows the t-SNE visualisation of \(f_{fg}\), \(f_{bg}\) and \(f_{uc}\) when we use auxiliary head and the corresponding loss functions (\(\mathcal{L}_{fg}\), \(\mathcal{L}_{bg}\) and \(\mathcal{L}_{compl}\)) to constrain the SID during training as well as when we do not use them.

We also tested the inference speed of our model, as shown in Table \ref{tab:fps}. On an NVIDIA GeForce RTX 4090, our model achieved an inference speed of 59.6 fps, fully meeting the needs for real-time segmentation.

\begin{table}[H]
\centering
\caption{Comparison of different model's inference speed using an NVIDIA GeForce RTX 4090.}
\label{tab:fps}
\begin{tabular}{cccc}
\toprule
Methods& TGANet& XBound-Former& Ours\\
\midrule
Inference& 32.7 fps& 61.3 fps& 59.6 fps\\
\bottomrule
\end{tabular}
\end{table}

Further, we verified the effects of using different backbone networks as the Encoder on the ISIC-2017 dataset. The original setup of our model uses ResNet-50 as the Encoder. In Table \ref{tab:encoder_performance}, we show the model performance when using various encoders, including VGG-19, ResNet-18, ResNet-34, ResNet-50, and Pyramid Vision Transformer (PVT). 

\begin{table}[H]
\centering
\small 
\caption{Performance of ConDSeg using different encoders.}
\label{tab:encoder_performance}
\begin{tabular}{@{}l@{}cccccc@{}} 
\toprule
Metrics~ & VGG-19 & ResNet-18 & ResNet-34 & ResNet-50 & PVT \\
\midrule
mIoU & 78.1 & 78.5 & 79.8 & 80.9 & 80.5 \\
mDSC & 85.9 & 86.5 & 88.0 & 88.3 & 88.3 \\
\bottomrule
\end{tabular}
\end{table}

To further validate the generalizability of our method, we extend it to multi-class 3D segmentation task. Specifically, we replace the original BCE loss with multi-class CE loss. Moreover, 3D images are converted into a series of 2D slice images following the experimental setup of TransUNet and SwinUNet. The experimental results on the Synapse dataset are shown as follows:

\begin{table}[H]
\centering
\small
\caption{Comparison of different methods on the Synapse dataset.}
\label{tab:synapse_results}
\begin{tabular}{@{}c@{}c@{}c@{}c@{}c@{}c@{}c@{}}
\toprule
Methods~~ & UNet~~ & TransUNet~~ & TransNorm~~ & MT-UNet~~ & SwinUNet~~ & Ours \\
\midrule
DSC $\uparrow$ & 76.9 & 77.5 & 78.4 & 78.6 & 79.1 & 80.2 \\
HD95 $\downarrow$ & 39.7 & 31.7 & 30.3 & 26.6 & 21.6 & 20.1 \\
\bottomrule
\end{tabular}
\end{table}

\section{Supplementary Visualizations}
\begin{figure*}[!t]
\centering
\includegraphics[width=\textwidth]{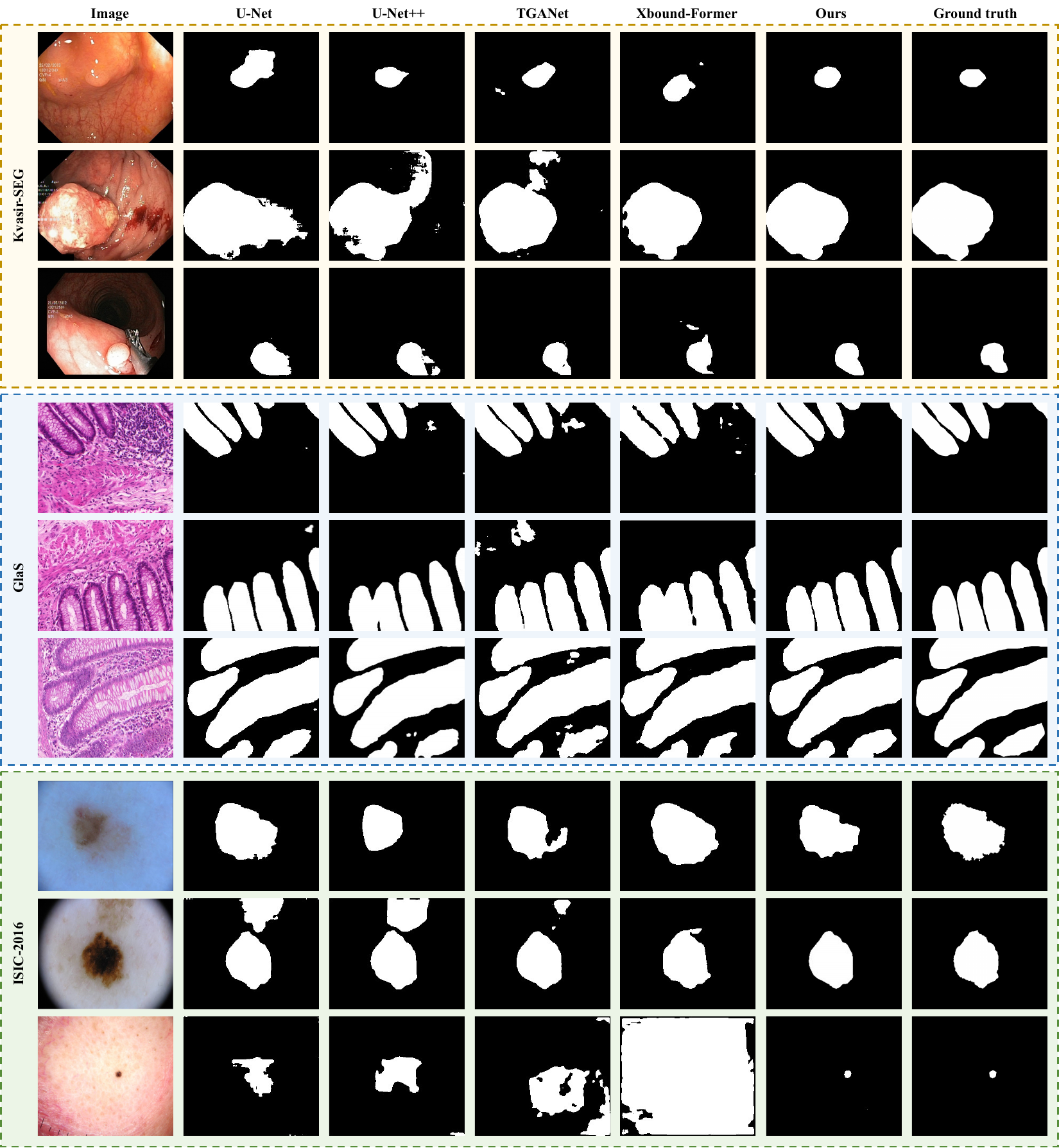}
\caption{Visualisation of segmentation results of different methods on datasets with different modalities.}
\label{fig:visualize}
\end{figure*}

We visualized the prediction results of our ConDSeg along with several other models, as shown in Figure \ref{fig:visualize}. Our model achieved prediction results that were closest to the ground truth across tasks of three modalities, significantly outperforming the other models.

\clearpage
\clearpage

\subsection{Supplementary Module Structure Diagrams}

\begin{figure}[H]
    \centering
    \vspace{60pt}
    \includegraphics[width=0.85\linewidth]{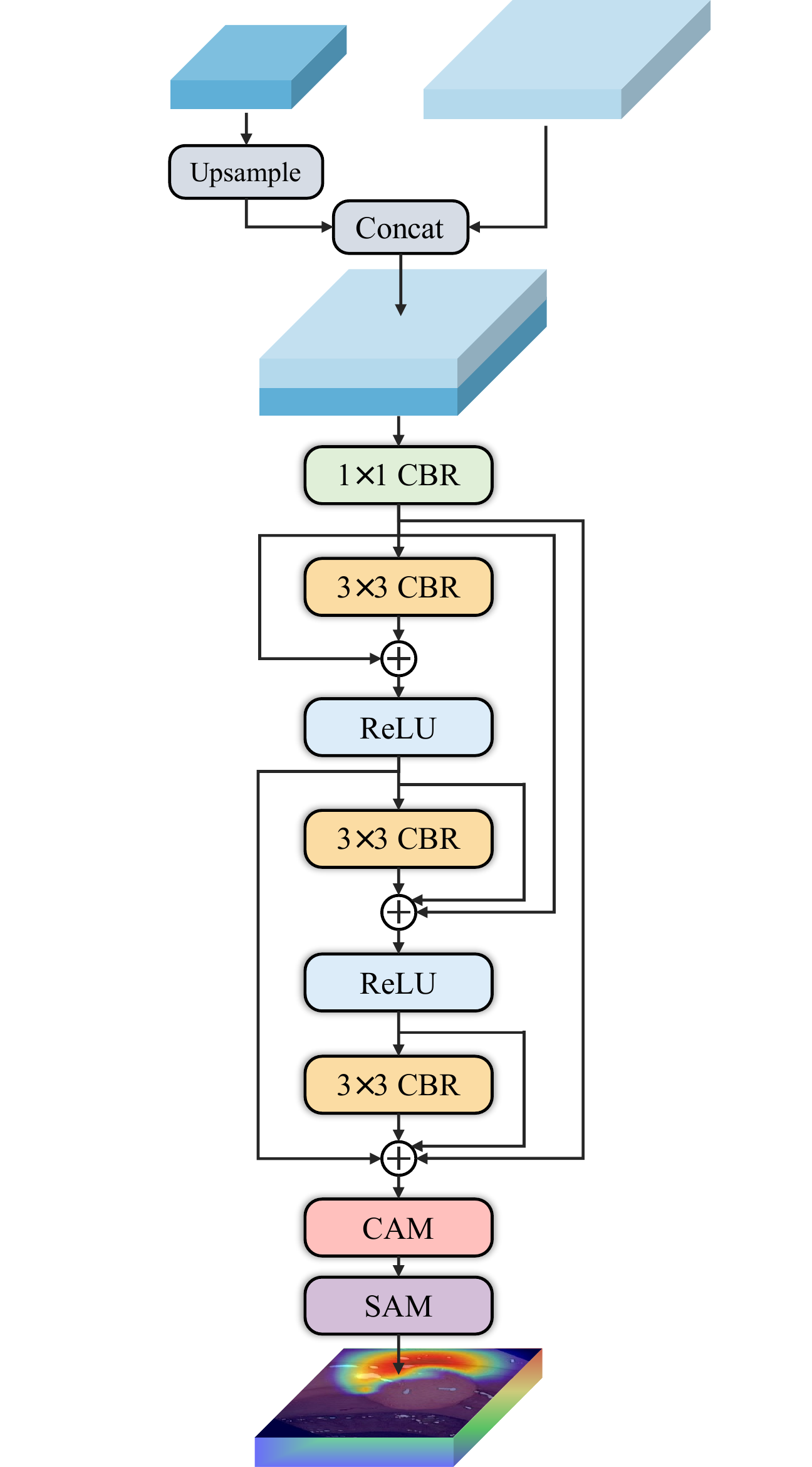}
    \caption{The structure of the proposed SA-Decoder. Here, CAM and SAM stand for channel attention module and space attention module, respectively.}
    \label{fig:sa-decoder}
    \vspace{80pt}
\end{figure}

\begin{figure}[H]
    \centering
    \includegraphics[width=0.85\linewidth]{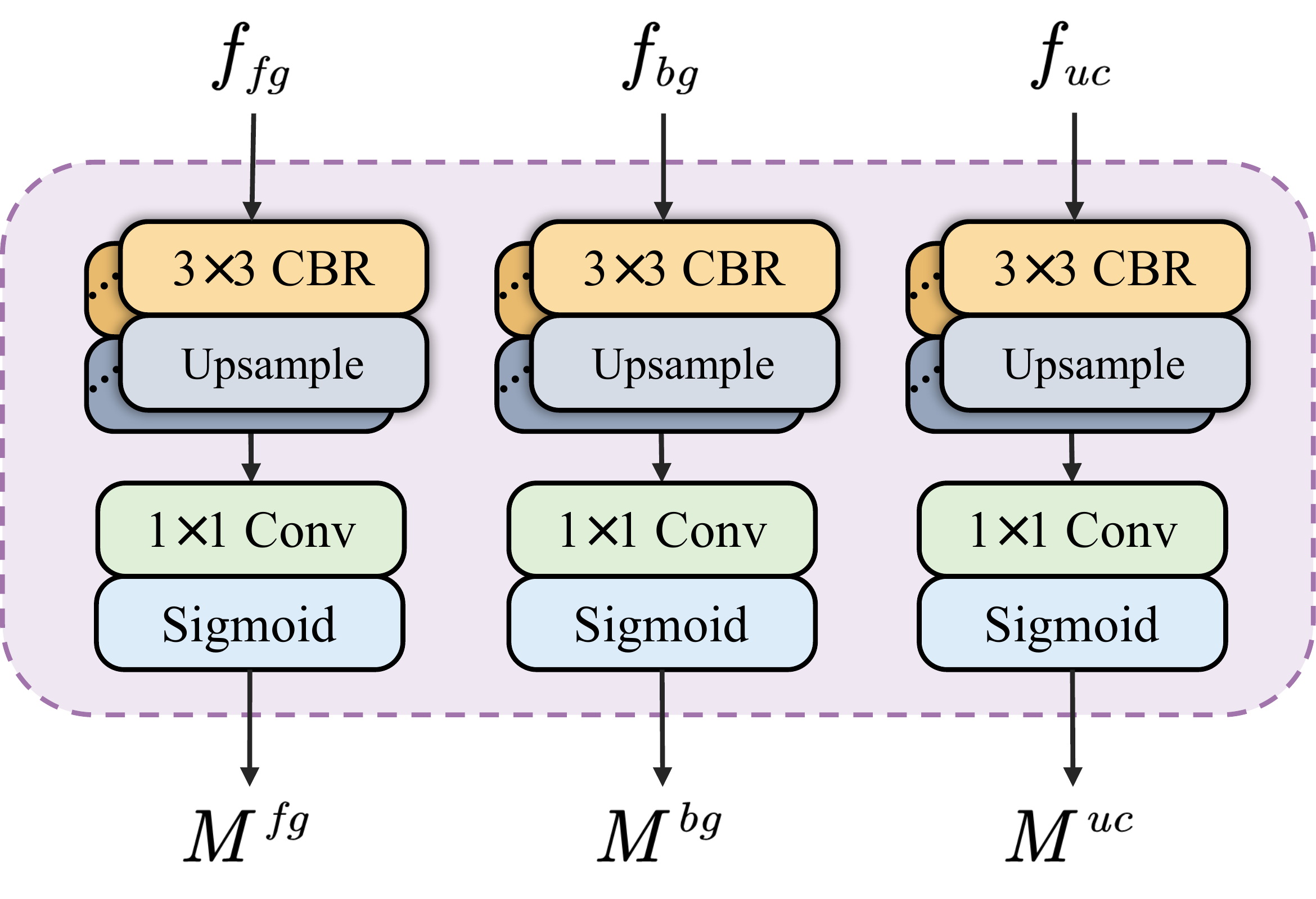}
    \caption{The structure of the auxiliary head in SID. Here, CAM and SAM stand for channel attention module and space attention module, respectively.}
    \label{fig:auxiliary}
\end{figure}

\begin{figure}[H]
    \centering
    \includegraphics[width=1\linewidth]{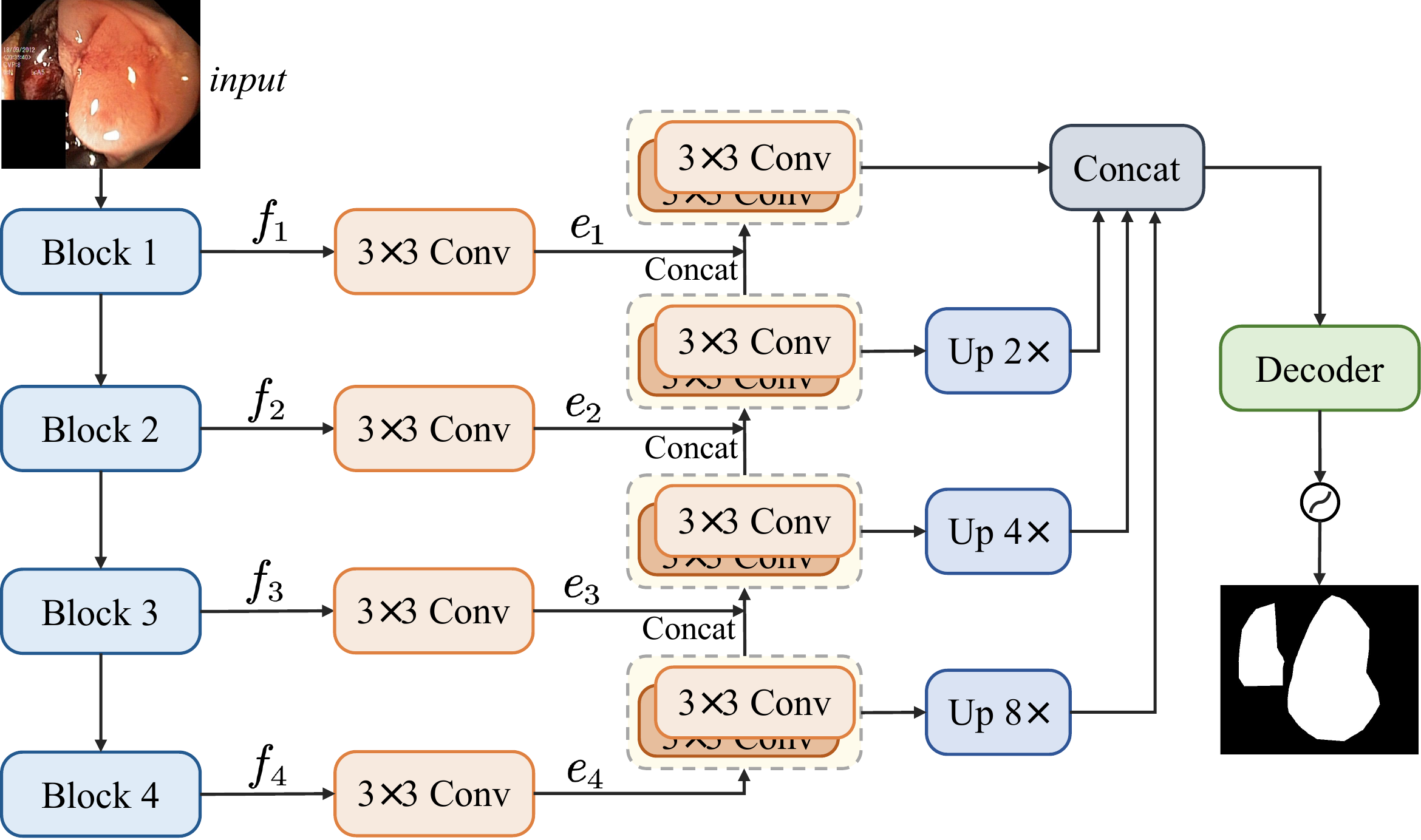}
    \caption{The structure of the baseline in the Ablation Study.}
    \label{fig:baseline}
\end{figure}

For some of the modules and networks mentioned in the main text, we provide detailed structures here. Figure \ref{fig:auxiliary}, Figure \ref{fig:sa-decoder}, and Figure \ref{fig:baseline} respectively showcase the structures of the auxiliary head, Size-Aware Decoder (SA-Decoder), and the baseline network.

\end{document}